\documentclass[final,1p,sort&compress,times,twocolumn]{elsarticle}
\pdfoutput=1

\usepackage[T1]{fontenc}
\usepackage[english]{babel}
\usepackage[intlimits]{amsmath}
\usepackage{array}
\usepackage{longtable}
\usepackage{calc}
\usepackage{multirow}
\usepackage{hhline}
\usepackage{ifthen}
\usepackage{parskip}
\usepackage{graphicx}
\usepackage{placeins}
\usepackage{listings}
\usepackage{braket}
\usepackage{breqn}
\usepackage{comment}
\usepackage{hyperref}
\usepackage{rotating}
\usepackage{todonotes}
\usepackage{upgreek}
\usepackage{here}
\usepackage{xcolor}
\usepackage{amssymb}

\newcounter{bla}

\journal{Computer Physics Communications}

\newcommand{\ambit}{\textsc{amb}{\footnotesize i}\textsc{t}}

\newfloat{lstfloat}{htbp}{lop}
\floatname{lstfloat}{Listing}

\begin{document}

\begin{frontmatter}

\title{\ambit: A program for high-precision relativistic atomic structure calculations}

\author{E. V. Kahl}
\author{J. C. Berengut\corref{author}}

\cortext[author] {Corresponding author.\\\textit{E-mail address:} julian.berengut@unsw.edu.au}
\address{School of Physics, University of New South Wales, Sydney NSW 2052, Australia}


\begin{abstract}
We present the \ambit\ software package for general atomic structure calculations. This software
implements particle-hole configuration interaction with many-body perturbation theory (CI+MBPT)
for fully relativistic calculations of atomic energy levels, electric- and magnetic-multipole transition
matrix elements, g-factors and isotope shifts. New numerical methods and modern high-performance 
computing techniques employed by this software allow for the calculation of open-shell systems with many
valence-electrons ($N \geq 5$) to a high degree of accuracy and in a highly computationally efficient 
manner.
\end{abstract}



\end{frontmatter}


{\bf PROGRAM SUMMARY}

\begin{small}
\noindent
{\em Program Title: \textnormal{\ambit}}                                          \\
{\em Licensing provisions: GPLv3}                                   \\
{\em Programming language: C++11}                                   \\
{\em Program available from: \url{https://github.com/drjuls/AMBiT}}            \\

{\em Nature of problem: Calculation of atomic/ionic spectra, including energy levels, electric and
magnetic multipole transition matrix elements, and isotope shifts.}\\
{\em Solution method: The program calculates energy levels and wavefunctions using either configuration
interaction (CI) only, or CI with many-body perturbation theory (CI+MBPT) in the 
{Brillouin-Wigner MBPT formalism}}.\\
{\em Restrictions: The program is not designed to treat highly-excited (Rydberg) states or continuum processes to a high degree of accuracy. }\\
   \\

\end{small}

\section{Introduction}

The combination of configuration interaction with many-body perturbation-theory (CI+MBPT) is one of the
workhorses of \textit{ab initio} high-precision atomic structure calculations and is known to provide
highly accurate results for many-electron atoms 
\cite{dzuba96a, berengut16a, torretti17a, kozlov15a}. Initially developed by Dzuba, Flambaum and Kozlov
\cite{dzuba96a} to calculate atomic energy spectra, CI+MBPT is also capable of providing other atomic
properties such as electronic transition data (e.g. electric- and magnetic-multipole transition 
matrix elements). 

The accuracy of CI+MBPT calculations lies in the ability to partition atomic structure calculations and
take advantage of the complementary strengths of CI and MBPT. The electrons are partitioned into 
``core'' electrons, which are treated as inert, and valence electrons, which display the dynamics of
interest. CI provides
a highly accurate treatment of valence-valence electron interactions (see, for example
\cite{fritzsche02a}), while MBPT treats the core-valence correlations in a computationally
efficient manner \cite{dzuba96a, berengut16a}. This combination of techniques
allows for the treatment of three \cite{dzuba96a}, four \cite{berengut06a,savukov15a} and five 
\cite{torretti17a, berengut11b} valence electrons, {with agreements with experimental spectra and
transition matrix elements to better than a few percent}.

Despite these advantages,
standard implementations, such as \texttt{CI+MBPT} of Kozlov \textit{et
al.} \cite{kozlov15a}, still require infeasibly large computational resources for $\gtrsim 4$ valence
electrons \cite{berengut16a}. Additionally, three-body MBPT corrections must be included to accurately 
treat systems with many valence electrons in this formalism \cite{dzuba96a,berengut08a, berengut11b}.
The number of 
these three-body MBPT diagrams grows extremely rapidly with MBPT basis size and can significantly
increase computation time \cite{berengut06a, berengut11b}, and are not included in the \texttt{CI+MBPT} code of Kozlov 
\textit{et al.}  \cite{kozlov15a}, for example.

Alternatively, pure configuration interaction (without the introduction of MBPT) is a frequently used 
approach in atomic structure software packages and is able to treat few-electron systems.
CI-only software packages include the \texttt{RELCI} 
package by {S. Fritzsche \textit{et al.}} \cite{fritzsche02a} (also part of the \texttt{GRASP}
package), 
the \texttt{PATOM} code of Bromley and co-workers \cite{jiang16a}. Of these packages, 
\texttt{PATOM} is only capable of calculating the spectra of one- and two-valence electron atoms, while 
\texttt{RELCI} only supports \textit{restricted} calculations for more than two valence electrons 
\cite{fritzsche02a}.

Another common method is multiconfigurational Dirac-Fock (MCDF), as implemented in the \texttt{GRASP} 
series of relativistic atomic structure packages \cite{jonsson13a} {and the MCDF code of Desclaux} 
\cite{desclaux75a}. 
The Flexible Atomic 
Code (\texttt{FAC}) software \cite{gu88a} has a similar rationale, calculating atomic spectra using 
Dirac-Fock and CI optimised for each configuration. \texttt{FAC} is commonly used for modeling atomic
processes in plasmas such as electron-impact ionization, excitation, and recombination processes.

Although pure CI and MCDF can provide a high degree of accuracy for few-electron
atoms, the number of many-electron configurations increases exponentially with the number of  
electrons \cite{dzuba10d}, making a direct solution with these techniques computationally infeasible for 
systems with $\gtrsim 4$ electrons \cite{berengut16a, kozlov15a}. The size of CI+MBPT calculations is
typically reduced by partitioning the electrons into core electrons, which are typically
treated with a self-consistent field method such as Dirac-Hartree-Fock, and ``valence'' electrons which
are directly included in the CI or MCDF procedure. While the partitioning reduces the computational
bottleneck from the total number of electrons to the number of valence electrons, it is still infeasible
to treat core-valence correlations without the introduction of MBPT 
\cite{berengut16a, dzuba96a}. Calculations by Kozlov \textit{et al.} \cite{kozlov15a} suggest that
CI+MBPT provides approximately an order-of-magnitude greater accuracy than pure CI due to the ability to 
treat core-valence interactions without dramatically increasing the size of the CI problem. 

Our atomic structure code, \ambit, has several advantages over existing CI+MBPT software, as well as
packages based around other numerical techniques. First, we implement three-body MBPT corrections, 
providing a significant
increase in accuracy for many-electron systems. Second, we can undertake the CI+MBPT procedure in either
the electron-only (as in most CI/CI+MBPT packages), or in the particle-hole formalism as presented 
in \cite{berengut16a}. This allows us to form open-shell (i.e. partially-filled) 
configurations from either all electrons or the
corresponding number of positively charged ``holes'' in an otherwise filled shell. For example, the
electron-only configuration $\ket{5d^{9} 6s}$ with the Fermi level below the 5d shell is equivalent to 
the particle-hole configuration $\ket{5d^{-1}6s}$ where the 5d shell is included in the core. The
electron-only and particle-hole approaches are formally equivalent at the CI level, but the 
particle-hole formalism can provide significantly more accurate MBPT corrections by reducing the
contribution of so-called ``subtraction diagrams'' \cite{berengut06a, dzuba05a}, which can seriously
degrade the accuracy of open-shell calculations. 

In addition to the standard core-valence MBPT corrections, we can also use MBPT to treat high-lying
valence-valence correlations \cite{berengut16a}. Valence-valence MBPT can
significantly reduce the size of the CI problem, especially for systems where the core and valence
electrons are separated by a relatively large energy gap (such as highly-charged ions). 
Additionally, we have developed a new addition to the standard CI+MBPT 
procedure, which we refer to as emu CI. This technique allows for a significant reduction in the
computational size of a CI+MBPT problem without significant reductions in accuracy \cite{geddes18a} and 
is discussed further in sections \ref{sec:SmallSide} and \ref{sec:comp}.

Our modifications to standard CI+MBPT allow for many-electron calculations with open-shell
configurations which would be either computationally infeasible or only possible with severely limited
accuracy in the standard framework. Additionally, \ambit\ makes extensive use of modern software
engineering techniques and parallel computing paradigms, allowing for highly efficient utilisation of
current high-performance computing (HPC) resources.

In this paper we give an overview of the \ambit\ code including the most important commands. We provide an example calculation of the energy level spectrum 
for Cr$^+$ -- a five valence electron system with an open d-shell. The spectrum of Cr$^+$ is very
well-characterised experimentally \cite{nistasd54}, but poses difficult computational and theoretical 
challenges to calculate with any degree of accuracy, making it an excellent test case for the 
extensive modifications to the CI+MBPT process implemented in \ambit.

\section{Overview of package}
The \ambit\ software consists of a single executable program, \texttt{ambit}, which 
carries out all aspects of the CI+MBPT calculations. The program does not require interactive input from
the user once invoked, with all input and usage options specified via either a single text file or
flags passed to the executable from the command line. These input options are interchangeable and accept 
the same input arguments and structure, {providing the flexibility} required to run \ambit\ 
through either small workstations and personal computers, or large-scale HPC clusters via batch-jobs.

\section{Installation}

\ambit\ is known to work on Linux and Apple OSX operating systems using GCC, Clang and Intel C++
compilers. The software may work on other operating systems or C++ compilers, but has not been tested.
The latest version of the software can be obtained from the \ambit\ GitHub repository at 
\url{https://github.com/drjuls/AMBiT}.

\subsection{Software dependencies}
The following libraries and tools are always necessary to compile \ambit:
\begin{itemize}
\item GSL - The GNU Scientific Library. 
\item The \textit{Boost} filesystem and system C++ libraries (boost\_filesystem and boost\_system).
\item Eigen v3 - C++ linear algebra package.
\item LAPACK and BLAS - linear algebra subroutines. Can be substituted for internal libraries in the 
Intel Math Kernel Library (MKL).
\item Google Sparsehash.
\end{itemize}

Additionally, the build system (outlined in section \ref{sec:scons}) requires the SCons build tool,
version 2.7 or higher of the Python programming language and the \texttt{pkg-config} Unix utility.

\ambit\ can also make use of MPI and OpenMP parallelism, as well as Intel MKL's built-in parallelism for
linear algebra subroutines. Currently, \ambit\ supports any conforming implementation of MPI, and can 
support OpenMP implementations from either GNU (GCC) or Intel. The use of these methods of parallelism 
must be explicitly enabled when compiling \ambit, to allow the software to support the maximum number of
platforms where not all of MPI, OpenMP and MKL are available (such as personal computers and
workstations).

\subsection{Compilation}
\label{sec:scons}

The \ambit\ build process is based around a modern build tool called SCons. Build options such as 
C++ compiler and compiler options, locations of required libraries, and which (if any) methods of 
parallelism to employ are specified in a plain-text file \texttt{config.ini}, an example of which is 
included with the source code as \texttt{config\_template.ini}. If no file named \texttt{config.ini} 
exists, then the build system will attempt to automatically create one with minimal build options.
Most build options can be left unspecified and the build system will attempt to automatically
infer sensible defaults, but these inferences can be explicitly overridden if required. {By default,
\ambit\ is compiled with dynamically linked libraries, but static linking can be forced by adding the
required compiler-specific flags, such as \texttt{-static} for GCC, to the \texttt{LINKFLAGS} field of
\texttt{config.ini} (see compiler-specific documentation for a more detailed treatment of compilation
options)}. 

{Additionally, the angular momentum data generated by \ambit\ (as outlined in section} 
\ref{sec:theory}) {is written to disk in a specified directory. This directory defaults to
\texttt{\$AMBITDIR/AngularData}, where \texttt{\$AMBITDIR} is the top-level \ambit\ directory, but can
be explicitly overridden via the \texttt{Angular data} input option in \texttt{config.ini}. This
directory contains the angular data from all runs of a particular \ambit\ installation, potentially
including calculations run by other users. Consequently, this directory must be created with the correct
file-system permissions in multi-user installations (such as HPC clusters)}.

Once the \texttt{config.ini} file has been filled as required, the software executable can be built by 
navigating to the top-level \ambit\ directory and issuing the \texttt{scons} command.

\section{Theory}
\label{sec:theory}
\subsection{CI + MBPT}
\label{sec:CI+MBPT}

Our CI+MBPT calculations consist of three conceptual stages. First, we treat the relatively inert
core electrons using the Dirac-Fock (DF) method, the relativistic generalisation of the Hartree-Fock
self-consistent field method. Second, we treat the valence electrons and holes with
configuration interaction using a set of B-spline many-electron basis-functions. Finally, the effects of
core-valence correlations and virtual core-excitations are included via many-body perturbation theory by
modifying the matrix elements used in the CI problem.

The full details of this process have been extensively discussed elsewhere (see, for example 
\cite{dzuba96a, berengut16a, berengut06a, torretti17a, johnson07a, geddes18a}), so we will only present
details relevant to our implementation of CI+MBPT. All calculations and formulae in this section are
presented in atomic units ($\hbar = e = m_e = 1$).

First, we perform a Dirac-Fock calculation, typically in 
the $V^N$, $V^{N-1}$ or $V^{N-M}$ approximations, where $N$ is the number of electrons and $M$ is the number of valence electrons. That is to say, we include either all $N$ electrons
of the ion, or some subset of them in the Dirac-Fock procedure. The choice of potential has
significant consequences for the convergence of the calculation, with a $V^N$ potential producing
``spectroscopic'' core orbitals, which are optimised for a particular configuration, while the $V^{N-M}$
potential (i.e. only including a subset of electrons in DF) potentially provides a better basis for the 
convergence of MBPT, by avoiding large contributions from so-called 
``subtraction diagrams'' \cite{dzuba05a}, which are discussed further below.

In either choice of potential, the resulting one-electron Dirac-Fock operator is (see, e.g.~\cite{johnson07a}):
\begin{align}
h_{\mathrm{DF}} = c\,\boldsymbol{\alpha} \cdot \mathbf{p} + (\beta - 1)c^2 - \frac{Z}{r} + 
V^{N_{\mathrm{DF}}}
\end{align}
where $\boldsymbol{\alpha}$ and $\beta$ are Dirac matrices. {We write the wavefunction as} 
\begin{eqnarray}
\psi(\mathrm{r}) = \frac{1}{r} \begin{pmatrix} f_{n \kappa}(r) ~\Omega_{\kappa,m}(\hat{r})\\
i g_{n \kappa}(r) ~\Omega_{-\kappa,m}(\hat{r}) \end{pmatrix}
\end{eqnarray}
{where $\kappa = (-1)^{j+l+1/2}(j+1/2)$ and $\Omega_{\kappa,m}$ are the usual spherical spinors. The
eigenvalue equation $h_{\mathrm{DF}}\,\psi_i = \epsilon_i \psi_i$ can be written in the form of coupled
ODEs}:
\begin{align}
\frac{df_i}{dr} &= -\frac{\kappa}{r} f_i(r) + \frac{1}{c}\left(\epsilon_i +\frac{Z}{r} - V^{N_{\mathrm{DF}}} + 2c^2 \right) g_i(r) \\
\frac{dg_i}{dr} &= -\frac{1}{c} \left(\epsilon_i +\frac{Z}{r} - V^{N_{\mathrm{DF}}} \right) f_i(r) + \frac{\kappa}{r} g_i(r)
\end{align}
{for each orbital $\psi_{i}$}.
{Numerical methods for solving these equations may be found in} \cite{johnson07a}.
{The resulting wavefunctions are used for orbitals in the Dirac-Fock core, while other basis orbitals
are constructed using B-splines as described below.}

At this stage, we may modify the Dirac-Fock operator to incorporate the effects of finite nuclear
size~
\cite{berengut08a}, nuclear mass-shift \cite{berengut06a, berengut11b}, and the Breit interaction 
{(including both Gaunt and retardation terms in the frequency-independent limit)} \cite{johnson07a}:
\begin{eqnarray}
B_{ij} = -\frac{1}{2r_{ij}}\left( \boldsymbol{\alpha}_i \cdot \boldsymbol{\alpha}_j + 
\left(\boldsymbol{\alpha}_i \cdot \boldsymbol{r}_{ij} \right) 
\left(\boldsymbol{\alpha}_j \cdot \boldsymbol{r}_{ij} \right)/r_{ij}^2
\right)
\end{eqnarray}
We may also include Lamb shift corrections, which are calculated via the radiative potential method 
originally developed by Flambaum and Ginges \cite{flambaum05a}. {The more recent formulation employed
in this code} includes the self-energy \cite{ginges16a} and vacuum polarisation \cite{ginges16b} 
corrections, collectively referred to in this paper as the QED corrections. {These corrections are 
propagated through the rest of the calculation via modification of the MBPT and radial CI (Slater) 
integrals, or the residual two-electron Coulomb operator in the case of the Breit interaction.}

We construct the remaining valence and virtual orbitals (pseudostates) as a linear combination of B-spline basis functions. {We
expand the large and small radial components, $f_{n\kappa}(r)$ and $g_{n\kappa}(r)$ of the virtual 
orbitals as linear combinations of two sets of B-splines $\{l_i \}$ and $\{ s_i \}$}:
\begin{eqnarray}
f_{n \kappa}(r) = \displaystyle\sum_{i} p_i l_i(r) \nonumber\\
g_{n \kappa}(r) = \displaystyle\sum_{i} p_i s_i(r)
\end{eqnarray}
{Each component of the wavefunction has the same set of expansion coefficients, which are obtained 
variationally by solving the generalised eigenvalue problem} \cite{johnson88a, beloy08a}:
\begin{eqnarray}
A \mathbf{p} = \varepsilon S \mathbf{p}
\end{eqnarray}
{where $A_{ij} = \bra{i}h_{\mathrm{DF}}\ket{j}$ is the matrix representation of the Dirac-Fock
operator in the B-spline basis, $S_{ij} = \left<i|j\right>$ is the overlap matrix}, 
$\ket{i} = \begin{pmatrix}l_i(r)\\s_i(r) \end{pmatrix}$ {are the B-spline basis functions, and 
$\varepsilon$ is the single-particle energy of the virtual orbital}.

{There is some freedom when choosing the exact for of the sets $\{l\}$ and $\{s\}$, as well as the
boundary conditions of the resulting B-spline basis functions. By default, \ambit\ uses the Dual 
Kinetic-Balance (DKB) splines developed in Ref.}~\cite{beloy08a} {due to their superior accuracy for 
atomic properties at small distances from the nucleus
and robustness against the effects of so-called ``spurious states''. However, alternative approaches can also be used, which in the terminology of \ambit\ are called ``Notre
Dame''} \cite{johnson88a} {and ``Vanderbilt''} \cite{fischer93a} {splines}. The resulting 
basis set is then ordered by energy and used for both CI and MBPT procedures.

{The one-particle basis functions are then used to construct a set of many-particle ``projections'', 
which are
(properly anti-symmetrised) configurations with definite angular momentum projection $m_j$ for every electron or hole. \ambit's
representation of projections is tightly coupled to their corresponding relativistic configurations --
projections corresponding to a particular relativistic configuration are represented as arrays of $2m_j$ 
angular momentum projections (which are always integer valued) of the orbitals in the configuration.}

{All projections corresponding to a relativistic configuration in the CI-space form a basis from
which to build many particle} Configuration State Functions (CSFs) $\{\ket{I}\}$. CSFs are
eigenfunctions of the $\hat{J}^2$ and $\hat{J}_z$ operators {and are formed as a linear combination
of projections:}
\begin{eqnarray}
\ket{I} = \displaystyle\sum_{n} c_n \ket{\mathrm{proj}_n}.
\end{eqnarray}
{We create $\ket{I}$ within the stretched state $M_J = J$, therefore only projections with
$\sum m_j = J$ are included in the expansion. The coefficients $\{c_n\}$ are determined
variationally by diagonalising the $\hat{J}^2$ operator in the projection basis:}
\begin{eqnarray}
\hat{J}^2 \ket{I} = J(J+1) \ket{I}
\end{eqnarray}
{To conserve memory, only the CSF expansion coefficients are stored in the Angular Data directory
(see section} \ref{sec:scons}) {and their matrix elements and integrals are calculated on the fly.}

The CI-space of CSFs are formed by taking electron excitations from a set of 
``leading configurations'' {(reference configurations) that are also used to determine which
three-body MBPT diagrams to include. Leading configurations can contain any number of valence electrons
and holes -- the only limit is the computational resources available when running the software.} We 
construct CI configurations and CSFs by taking excitations from these leading configurations up to some
maximum principal quantum number, $n$, and orbital angular momentum, $l$. {These limits are
represented in the software and throughout the rest of this paper using a shorthand representation of
called a ``basis string''. For example, we can specify orbitals with $0 \leq l \leq 3$ and $n \leq 10$ 
for each partial wave via the string 10spdf, or s- and p-orbitals with $n \leq 10$ and d-orbitals with 
$n \leq 7$ with the string 10sp7d. Further details of CI input and output are presented in section}
\ref{sec:comp}.

The projections and CSF expansion coefficients for each configuration with the same number of electrons,
$J^{\pi}$ symmetry and projection $M_J$ are stored to disk. This allows the initial cost of 
diagonalising the $\hat{J}^2$ operator to be amortised across all calculations with the same angular 
components, dramatically reducing the overall computational cost.

\ambit\ can construct CSFs to use in CI calculations with an arbitrary number of electron excitations, 
but finite computational resources usually limit the CI basis to single- (often referred to as CIS) or
single- and double-excitations (CISD). However, important triple or quadruple excitations should also be
included. The CSFs can also include valence-holes in otherwise filled shells, which can lie between the
Fermi level of the system and some minimum $n$ and $l$, the latter of which is referred to as the
``frozen core''.

The atomic level wavefunctions $\Psi$ for a given total angular momentum and parity $J^{\pi}$ are then 
constructed as a linear combination of CSFs $\ket{I}$:
\begin{align}
\label{eq:Psi_CI}
\Psi = \displaystyle\sum_{I \in P} C_{I}\ket{I}
\end{align}
where $P$ is the subspace of configurations included in CI and the coefficients $C_{I}$ are obtained
from the matrix eigenvalue problem of the CI Hamiltonian: 
\begin{align}
\displaystyle\sum_J H_{IJ}C_J = EC_{I}
\end{align}
In the particle-hole formalism, the CI Hamiltonian is \cite{berengut16a}:
\begin{align}
\hat{H} =  \displaystyle\sum_{i} c\,\boldsymbol{\alpha} \cdot \mathbf{p} + (\beta -
1)c^2 + \frac{Ze_i}{r_i} - e_{i}V^{N_{\mathrm{core}}} +
\displaystyle\sum_{i < j} \frac{e_i e_j}{|\mathbf{r}_i - \mathbf{r}_j|}
\end{align}
where $e_i = -1$ for valence electron states and $+1$ for holes. It is important to note that the 
one-body potential $V^{N_{\mathrm{core}}}$ in the CI Hamiltonian only includes contributions from the 
core electrons, since valence-valence correlations are included directly via the two-body Coulomb 
operator.

The size of the CI matrix grows extremely rapidly as additional orbitals are included, so it is 
computationally infeasible to include core-valence correlations or core excitations directly in the CI
procedure. Instead, we treat these interactions as a small perturbation and include
their contributions to the total energy in second-order MBPT by modifying the CI matrix elements
\cite{dzuba96a, berengut06a}. The final matrix eigenvalue problem for the CI+MBPT technique is then:
\begin{align}
\label{eq:MBPT}
\displaystyle\sum_{J \in P}\left(H_{IJ} + 
\displaystyle\sum_{M \in Q} \frac{\bra{I} \hat{H} \ket{M} \bra{M}\hat{H} \ket{J}}{E - E_M} \right)C_J
= E C_I
\end{align}
where the subspace $Q$ includes all orbitals not in the CI procedure and is complementary to $P$.
For computational efficiency, we do not directly modify the CI matrix elements as suggested by equation
(\ref{eq:MBPT}), due to the large number of configurations in $Q$. Instead, we modify the radial 
integrals via the Slater-Condon rules for calculating matrix elements (see ref \cite{kozlov99b} for a 
formal discussion of this process).

The subspace $Q$ is formally infinite, but we only include corrections from a finite, truncated subset 
of orbitals in MBPT. {In \ambit, the $P$ and $Q$ subspaces are divided at the level of orbitals. The
CI space $P$ includes any configuration with all single-particle orbitals drawn from the valence basis
or holes outside the frozen core; these in turn are defined by two basis strings, described further in
section} \ref{sec:comp}. {Similarly, the MBPT space $Q$ is bounded by a separate orbital limit, also
expressed as a basis string (e.g. 30spdfg for all orbitals with $n \leq 30$ and $l \leq 4$). The input
options controlling this are discussed further in section} \ref{sec:comp}.

{Consequently, an excitation is only included in MBPT if at least one of the orbitals involved are
not included in the CI space $P$. This prevents double-counting of configurations and ensures that
diagrams are independent of the number of electron- and hole-excitations in CI.}

\begin{figure}
\label{fig:one_body_MBPT}
\includegraphics[width=0.7\textwidth]{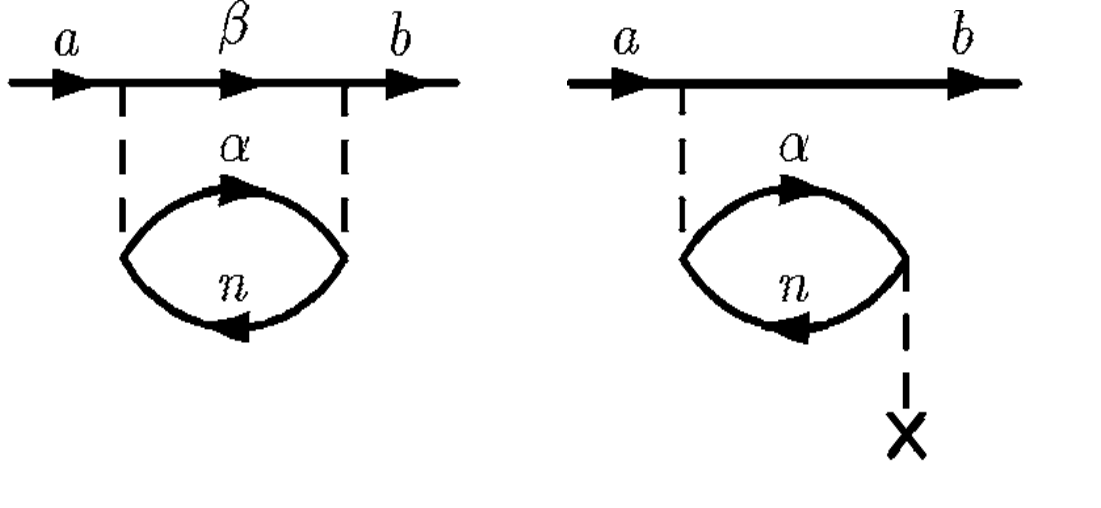}
\centering
\caption{Some Goldstone diagrams representing a one-body core-valence correlation (left) and 
one-body subtraction diagram (right). Lines running left to right represent electrons, while lines 
running right to left are holes. $\ket{a}$ and $\ket{b}$ are valence orbitals, 
$\ket{\alpha}$ and $\ket{\beta}$ are virtual, and $\ket{n}$ is a hole in the core~\cite{berengut06a}.}
\end{figure}

The number of terms in the MBPT corrections grows rapidly but the diagrammatic
technique~\cite{dzuba96a} greatly simplifies the calculation of these terms. In
this formalism, each contribution to the MBPT expansion is represented by a Goldstone diagram, with the
number of external lines corresponding to the number of valence electrons included in the interaction
\cite{berengut06a}. Figure~\ref{fig:one_body_MBPT} shows an example of a one-body MBPT diagram describing
the self-energy correction arising from core-valence interactions (left) and a subtraction diagram
involving an interaction with an external field (right) \cite{berengut06a}. These subtraction diagrams
enter the MBPT expansions with a negative sign and increase in magnitude with 
\mbox{$V^{N_{\mathrm{core}}} - V^{N_{\mathrm{DF}}}$} \cite{dzuba05a}. {Explicit formulas for one-
and two-body core-valence diagrams implemented in \ambit\ can be found in Ref.}~\cite{berengut06a}.

Subtraction diagrams are partially cancelled out by some two- and three-body diagrams in the MBPT 
expansion \cite{berengut11b}, necessitating the systematic inclusion of all one-, two- and three-body
MBPT diagrams in the CI+MBPT procedure to ensure accurate spectra. Even given this cancellation,
subtraction diagrams can grow large enough to be non-perturbative in open-shell systems, which can
significantly impact the accuracy of the resulting spectra \cite{berengut11b}. Consequently, there is a 
tradeoff between the more ``spectroscopic'' orbitals produced by calculations in a $V^N$ potential and 
potentially large subtraction diagrams when $V^{N_{\mathrm{DF}}} \neq V^{N_{\mathrm{core}}}$; the 
optimal choice will depend on the specifics of the target system. {This is not a hard constraint
though -- the formulation of MBPT used in \ambit\ can, in principle, treat systems with any number of
valence electrons or holes subject to available computational resources.}

An additional complexity is that the energy denominators of (\ref{eq:MBPT}) include the energy 
eigenvalue $E$ in the Brillouin-Wigner perturbation theory formalism. In practice we approximate the 
energy denominators using the valence orbital energies~\cite{berengut06a}. See 
Refs.~\cite{dzuba96a,kozlov99b} for further discussion of this subtle point.
Finally, the diagrammatic technique allows us to eliminate terms
corresponding to unlinked diagrams, as they represent valence electron interactions not included in MBPT
\cite{dzuba96a}. This greatly reduces the computational expense of including MBPT corrections.

\begin{figure}
\includegraphics[width = 0.7\textwidth]{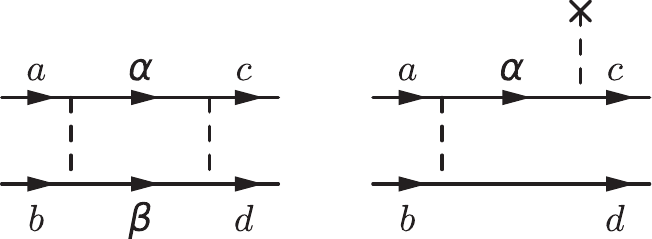}
\centering
\caption{Two-body valence-valence diagram (left) and valence-valence subtraction diagram (right). 
External lines $a, b, c, d$ correspond to valence or hole 
orbitals in the CI subspace $P$, while the interior lines $\alpha$ and $\beta$ are virtual electron 
orbitals, at least one of which must not be valence for the intermediate state to be in $Q$~\cite{berengut16a}.}
\label{fig:valence_MBPT}
\end{figure}

In addition to the standard core-valence MBPT, \ambit\ can also include MBPT corrections to
valence-valence integrals, as introduced in \cite{berengut16a}. In this approach, the MBPT expansion in
equation (\ref{eq:MBPT}) includes additional diagrams representing correlations between highly-excited 
valence states (i.e. outside the upper-bounds of the CI-space $P$), as shown in figure
\ref{fig:valence_MBPT}. 

Valence-valence MBPT is significantly computationally cheaper than including the
orbitals directly in the full CI subspace. However, as with all MBPT techniques, care must be taken to 
ensure that there are no non-perturbative diagrams in the MBPT expansion. Specifically, including 
orbitals which are far from spectroscopic (such as orbitals with high orbital angular momentum) or 
orbitals which are close in energy to those in $P$ can produce non-perturbative diagrams with small
energy-denominators. These diagrams can significantly reduce the accuracy of CI+MBPT calculations and
are easy to inadvertently include in the MBPT expansion, so this approach should only be used with a
carefully constructed MBPT subspace $Q$. 

The energies for each calculated eigenstate are presented by \ambit\ in ascending order of energy, 
grouped 
by total angular momentum $J$ and parity $\pi$. The solutions also contain the CI expansion coefficients
and Land\'{e} g-factors, to aid in identifying levels. It is important to note that the
absolute energies do not represent ionisation energies or any other physically meaningful quantity: the atom is effectively in a box due to the finite extent of the basis orbitals. Rather only the 
\emph{relative} energies of the eigenstates (and the resulting atomic spectrum) represent physically
meaningful quantities.

Finally, the resulting CI+MBPT wavefunctions are used to calculate transition matrix elements for
{electric and magnetic multipole operators, which we refer to as ``external field'' operators}, 
as well as hyperfine dipole and quadrupole operators. {\ambit\ can calculate either reduced matrix
elements $T$:}
\begin{eqnarray}
T_{if} = \bra{f} \hat{O} \ket{i}
\end{eqnarray}
{for some operator $\hat{O}$, initial state $\ket{i}$ and final state $\ket{f}$, or line-strengths:}
\begin{eqnarray}
S_{if} = \left|T_{if}\right|^2 
\end{eqnarray}
Transition matrix element calculations may additionally include frequency-dependent random-phase 
approximation (RPA) corrections \cite{johnson80a, dzuba84a, dzuba86a}; {detailed equations
implemented in \ambit\ are presented in Ref.}~\cite{dzuba18a}.

\subsection{Emu CI}
\label{sec:SmallSide}

The CI method outlined in section \ref{sec:CI+MBPT} relies on constructing and diagonalising the Hamiltonian matrix over 
a set of many-electron CSFs. The number of CSFs, and consequently the size of the CI matrix, scales 
exponentially with the number of electrons included in the CI problem subspace, resulting in
prohibitively large CI matrices for systems with more than three valence electrons. Additionally, CI is 
slowly converging even for relatively simple systems with few valence electrons \cite{bromley07a}, 
making saturation in open-shell systems infeasible with current computational methods. 

\begin{figure}
\center
\includegraphics[width=0.5\textwidth]{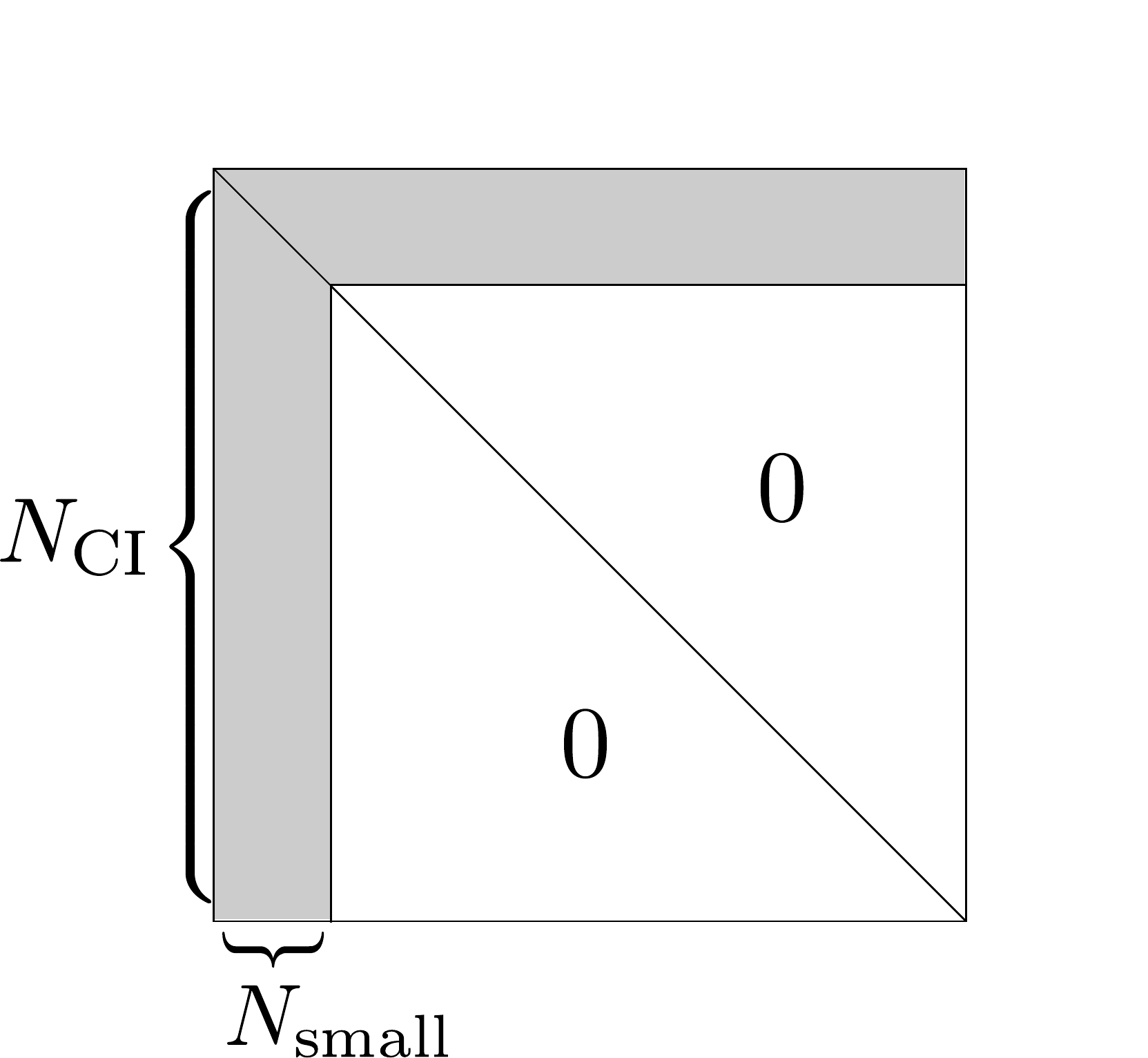}
\caption{Structure of the CI matrix under the emu CI approximation. 
The $N_{\mathrm{CI}} \times N_{\mathrm{small}}$ nonzero off-diagonal terms are shaded in light grey,
terms neglected in the approximation are shown in white}
\label{fig:smallside}
\end{figure}

\ambit\ implements a new approach that greatly reduces the computational difficulties associated with 
CI, which we
refer to as emu CI~\cite{geddes18a} (as the structure of the resulting CI matrix resembles an emu's
footprint). This approach is especially well-suited to the common case where we are 
only interested in calculating a few of the lowest-lying energy levels and allows for the use of 
significantly larger CI basis sizes than would otherwise be possible. A schematic representation of this 
approach is shown in figure \ref{fig:smallside}. 

Emu CI relies on the fact that the CI expansion (\ref{eq:Psi_CI}) is dominated by relatively few large
contributions from 
off-diagonal CI matrix elements. Other CSFs contribute less strongly, and so interactions between these may be neglected.
{The shaded region of the matrix shown in figure} \ref{fig:smallside} {is formed as the Cartesian
product of the $N_{\mathrm{CI}}$ CSFs in the CI-space $P$, which we refer to as the ``large side'' of the matrix, and a smaller set of 
$N_{\mathrm{small}} < N_{\mathrm{CI}}$ CSFs, which we refer to as the ``small side'' of the matrix.
The small side contains a subset of CSFs that make the largest contribution to the CI expansion.}
Perturbation theory estimates performed in \cite{dzuba17a} show that the 
remaining off-diagonal terms, shown in white, produce a negligible contribution for the small number of states of interest, and can be set to zero without serious loss of accuracy.

{The small-side CSFs are formed by allowing electron and/or hole-excitations from a set of leading 
configurations (which is not necessarily the same as used when forming the main CI-space) up to some
maximum principal quantum number $n$ and orbital angular momentum $l$. This limit is specified using the
same format of basis string used when forming the main CI-space. Finally, the small-side can be include
an arbitrary number of electron- and hole-excitations (which also do not have to be the same as in the 
main CI-space).}

We can then construct the CI matrix such that the significant off-diagonal terms are grouped
together in a block, producing the structure shown in figure \ref{fig:smallside}. These elements are
further sorted such that the {configurations with the largest number of corresponding projections
appear first in the matrix} to
provide better performance when constructing and diagonalising the matrix in parallel.

The dramatically reduced number of non-zero elements in the emu CI matrix compared to standard CI 
significantly reduces the computational resources required to obtain accurate atomic spectra. 
Recent calculations of the spectra of neutral tantalum and dubnium \cite{geddes18a} shows that emu CI
is capable of producing highly accurate atomic spectra for five-electron systems despite its relatively
small resource usage. Spectra from these calculations were within $\sim\!10$\% of experimental values, and
convergence tests showed that the CI expansion was close to saturation. Similarly, applying this
technique to the Cr$^{+}$ calculations  presented in this paper reduces the number of non-zero elements 
in the effective CI by a large factor. Even larger reductions in matrix size were used in
\cite{geddes18a}. The full matrices 
would be far to large to store in memory, even on modern high-performance computing clusters.
Emu CI, combined with modern parallel programming techniques enables the use of extremely large CI bases, even for challenging open-shell systems with strong correlations.

\section{Description of program components}
\label{sec:comp}

The aspects of the CI+MBPT calculation presented in section \ref{sec:theory} are controlled by a set of 
input options supplied to the software at run-time. This section contains a list of the most commonly
used options when performing a full CI+MBPT calculation; a more comprehensive list can be found in the
\ambit\ user guide supplied with the software.

Input options are specified as either command-line arguments passed to the \texttt{ambit} executable, or
contained in a plain-text input file. The input file must end with the suffix \texttt{.input} and be
passed as a command line argument when calling the \texttt{ambit} executable. Although both input
methods are equivalent and can be used interchangeably, we will assume that the input options are 
specified in an input file in this section. 

\begin{figure}
\includegraphics[width=0.8\textwidth]{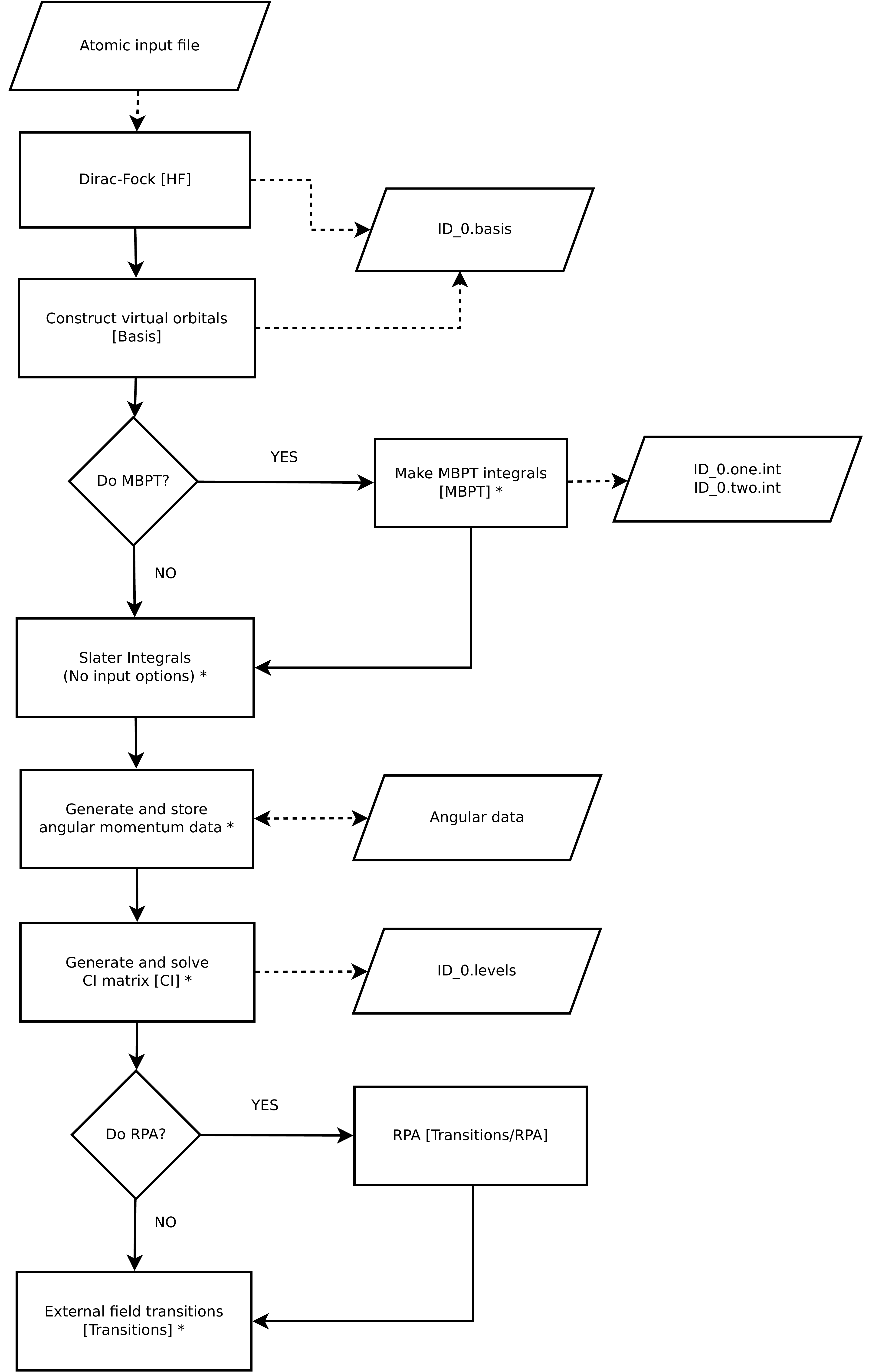}
\caption{{Flow chart representing control and program flow of \ambit. Rectangular blocks represent
physical sections of the calculation outlined in section} \ref{sec:theory}, {with the corresponding
input file section (if any) listed in square brackets. Program sections labeled with an asterisk
\texttt{*} are parallelised with MPI, OpenMP or both. Solid arrows represent program control flow, while
dashed arrows represent input/output files consumed/created by \ambit.}}
\label{fig:program_flow}
\end{figure}

Input options are organised into a series of \textit{sections} loosely corresponding to the physical
sections of the calculation outlined in section \ref{sec:theory}. {Figure} \ref{fig:program_flow} 
{provides a high-level overview of program flow, as well as the correspondence between input sections
and physical methods.}

Each section begins with a
\textit{header} in square brackets and continues until the next header (or end-of-file) and can contain
any number of subsections of the form: 

\begin{verbatim}
[Section]
Argument
[Section/Subsection]
Argument
\end{verbatim}

Equivalently, sections and sub-sections may be included in a single-line argument with the format:

\begin{verbatim}
Section/Argument
Section/Subsection/Argument
\end{verbatim}

The latter format can also be supplied as command line arguments when invoking \ambit. These sections
can occur in any order in the input.

Options in CamelCase (i.e. capitalisation denotes separate words) accept a value
with the syntax \texttt{Option = value}. Options consisting of lower-case words separated by 
dashes (e.g. \texttt{{-}{-}check-sizes}) are flags used to toggle behaviour on or off and do not accept 
values. Input options not specified as mandatory may be left out of the input and will default to values 
given in square brackets next to the option.

\ambit\ will print all supplied arguments, as well as the current version, at the top of each 
calculation's output; this ensures that any individual calculation can be exactly repeated 
using only information presented in the software output. In addition, the software saves the set of 
basis orbitals, one- and two-body MBPT integrals and atomic wavefunctions obtained by CI as binary files 
in the current working directory (i.e. the location from which the \texttt{ambit} executable is 
launched), allowing calculations to be cancelled and resumed, as well as for results from previous 
calculations to be reused in future calculations.

Most options specifying orbitals or configurations take non-relativistic inputs, which are automatically
converted into the corresponding set of relativistic orbitals or configurations for use in the rest of
the calculation. 

\subsection{General options}
\label{sec:general_options}
The following general options do not fall under any named section of the input and must be specified 
at the beginning of the input file (i.e. before any named sections are introduced):

\textbf{ID} An identifier for the calculation, used to label the binary files generated during the
calculation. (Mandatory.)

\textbf{Z} Atomic number of the atomic system. (Mandatory.)

\textbf{-s{1,2,3}} Specifies which of the one-, two- or three-body MBPT diagrams to include in the
calculation. Any combination
of $1,2$ or $3$ may follow the \texttt{-s}, so e.g. \texttt{-s1} will only calculate one-body MBPT
corrections, while \texttt{-s123} will calculate one-, two- and three-body corrections. No MBPT
corrections will be calculated if no variations of this flag are set.

\textbf{{-}{-}no-new-mbpt} Do not calculate any new MBPT integrals. Any integrals saved in 
\texttt{.int} files will still be read, however integrals that have been requested, but are not present in 
the \texttt{.int} files, will not be calculated. 

\textbf{NuclearInverseMass} Inverse of the nuclear mass in atomic units: 
$1/M_{\mathrm{nuc}}$. Used to calculate isotopic mass-shift.

\textbf{NuclearRadius} Specifies the radius parameter of a Fermi distribution for nuclear charge. This 
value is used to compute the nuclear RMS radius, which is automatically used to
calculate field-shift contributions.
If either \texttt{NuclearRadius = 0} or \texttt{NuclearThickness = 0} are set then the nucleus is
treated as a point charge.

\textbf{NuclearThickness} Specifies the density parameter of a Fermi distribution for nuclear charge. 
If either \texttt{NuclearRadius = 0} or \texttt{NuclearThickness = 0} are set then the nucleus is
treated as a point charge.

\textbf{{-}{-}check-sizes} Calculates and prints the number of MBPT
and Coulomb integrals, as well as the number of CSFs per symmetry $J^{\pi}$ (i.e. the size of the CI
matrices) without calculating any integrals or energy levels. Used to obtain a rough estimate of the 
size of the desired calculation before running it in full. This option will also calculate any angular
data (if required), which can be computationally intensive and often requires \ambit\ to be run in
parallel to complete quickly.

\subsection{[Lattice]}
The \texttt{[Lattice]} input section specifies the parameters of the numerical lattice on which the
basis functions and
solutions are calculated. The lattice uses exponential spacing close to the nucleus and transitions to
linear spacing at larger radii. Points in ``real-space'', $r$, are related to the uniformly-spaced set 
of lattice-points $x$ by:
\begin{align}
x = r + \beta\ln(r/r_0)
\end{align}
where $r_0$ is the starting point of the lattice in atomic units (corresponding to
\texttt{Lattice/StartPoint}) and $\beta$ controls the distance at which the lattice transitions from
exponential to linear spacing.

[Lattice] contains {the following options}:

\textbf{NumPoints}[1000] Number of points in the lattice.

\textbf{StartPoint}[1.0e-6] Starting point of the lattice in atomic units. Must be nonzero.

\textbf{EndPoint}[50.0] Last point in the lattice (in real-space).

\subsection{[HF] - Dirac-Hartree-Fock}

\label{sec:comp_HF}

The software allows for considerable freedom in specifying the Dirac-Hartree-Fock component of a
calculation. In general, any number of electrons can be included in the procedure in any desired (valid)
configuration. The choice of Dirac-Fock potential $V^{N_{\mathrm{DF}}}$ as described in section~\ref{sec:theory} can severely impact the accuracy of the resulting spectra; with
a potential $V^{N_{\mathrm{DF}}} = V^{N}$ producing ``spectroscopic'' core orbitals (which benefit CI
convergence) but potentially introducing large subtraction diagrams for open-shell systems (which can
reduce MBPT accuracy).

Additionally, the Dirac-Hartree-Fock section of the calculation allows for the inclusion of a number of
``decorators'', such as the Breit interaction or QED corrections, which modify the Dirac-Fock operator.
Decorators introduced in this step of the calculation persist for the lifetime of a particular run of
\ambit, and are also incorporated in the CI (Slater) and MBPT integrals. These decorators are controlled
by flags passed to the [HF] section of the input file.

The [HF] section accepts the following arguments and flags:

\textbf{N} The number of electrons to include in the Dirac-Fock procedure. (Mandatory.)

\textbf{Configuration} String specifying the nonrelativistic configuration to be included in the
Dirac-Fock calculations (can include valence as well as core electrons). The Fermi level can be specified
with a colon (`:'), orbitals above which will be included in the CI and MBPT valence space $P$. If no 
Fermi level is specified it will default to immediately above the highest supplied orbital in this 
argument. Additionally, Valence holes can only appear in shells with energy below the Fermi level.

As an example, the string \texttt{HF/Configuration = '1s2 2s2 2p6~:~3s1'} will include the configuration
1s$^2$ 2s$^2$ 2p$^6$ 3s, with 1s, 2s and 2p shells below and 3s above the Fermi level.
(Mandatory; configuration must have \texttt{HF/N} total electrons.)

\textbf{{-}{-}breit} Decorator including effects of the Breit interaction.

\textbf{{-}{-}sms} Decorator for Specific Mass Shift operator for isotope shift.

\textbf{{-}{-}nms} Decorator for Normal Mass Shift operator for isotope shift.

\subsection{[HF/QED] - QED corrections}

The decorators incorporating the Lamb shift corrections to the atomic spectra are
controlled by arguments in the [HF/QED] section of the input file, which accepts the following 
arguments:

\textbf{{-}{-}uehling} Decorator for the Uehling (vacuum polarisation) correction.

\textbf{{-}{-}self-energy} Decorator for the self-energy correction.

\subsection{[Basis] - CI+MBPT basis functions}

As outlined in Section \ref{sec:CI+MBPT}, {single-particle basis orbitals are obtained by} 
diagonalising a large set of B-Splines against the Dirac-Fock operator. The B-Spline
basis includes orbitals up to some maximum principal quantum number and orbital angular momentum as
determined by the \texttt{Basis/ValenceBasis} or \texttt{MBPT/Basis} input arguments, whichever is
higher. This input section also determines the extent of the CI space $P$. The CI basis will be
constructed using electron orbitals between the Fermi level and the upper limit set by
\texttt{Basis/ValenceBasis} and hole orbitals between the lower limit set by
\texttt{Basis/FrozenCore} and the Fermi level.

Once the basis has been generated, it is saved to the binary file \texttt{<ID>_0.basis} (where 
<ID> is given by the corresponding input option in section \ref{sec:general_options}) in the current 
working directory. The software checks for the existence of this file in the current working directory 
and will reuse any pre-calculated orbitals rather than generate a new set for each calculation.

The [Basis] section of the input file accepts the following arguments:

\textbf{ValenceBasis} String specifying the maximum principal quantum number, $n$, and orbital angular
momentum, $l$, of
the orbitals included in the CI space $P$. The electron orbitals included in CI will therefore run from
the Fermi level up to the limit specified by this option. For example, 
\texttt{Basis/ValenceBasis = 10spdf} will include orbitals with $0 \leq l \leq 3$ and $n \leq 10$ for
each partial wave. It is also possible to specify different values of $n$ for each partial wave, so, for
example, \texttt{Basis/ValenceBasis = 10sp7d} will include s- and p-orbitals with $n \leq 10$ and
d-orbitals with $n \leq 7$. (Mandatory.)

\textbf{FrozenCore} Sets an upper limit to the frozen core, in which the shells cannot have holes. This argument accepts the same
input syntax as \texttt{Basis/ValenceBasis}, so \texttt{Basis/FrozenCore = 4sp3d} will not allow holes in
s- and p-shells with $n \leq 4$ and d-shells with $n \leq 3$. If this option is omitted then the
frozen-core will be set at the Fermi level (i.e. no holes will be allowed in CI).

\textbf{{-}{-}bspline-basis} Construct the basis functions from B-Splines.

\subsection{[Basis/BSpline] - B-Spline options}

While it is generally fine to use the default values for the low-level, mathematical properties of 
B-Splines used to generate the basis functions, it is also possible to directly set these properties with
the arguments in this section. See refs. \cite{johnson88a} and \cite{beloy08a} for a full discussion of
the use of B-Splines as basis functions.

\textbf{N} [40] Number of splines per basis function.

\textbf{K} [7] Order of splines (splines are polynomials of degree $k - 1$ and are nonzero across $k$
intervals). Default value is 7, but this will be automatically adjusted so $k \geq l_{\mathrm{max} + 3}$.

\textbf{Rmax} [50.0] Maximum radius of the B-Splines. This should be chosen so that valence orbitals are not seriously impacted by the truncation at finite radius.

\textbf{SplineType} [Reno] Specifies the boundary conditions used to eliminate spurious states from the
basis. Options are \texttt{Reno} (the default) \cite{beloy08a}, \texttt{NotreDame} \cite{johnson88a}
or \texttt{Vanderbilt} \cite{fischer93a}.

\subsection{[CI] - Configuration Interaction}
\label{sec:comp_CI}

The CI matrix eigenvalue problem is solved by one of two methods, depending on the size of the 
matrix and number of required solutions. For small matrices with less than $N = 200$ CSFs, we simply 
solve the eigenproblem directly using an implementation of the \texttt{LAPACK} linear algebra library 
such as Intel's \texttt{MKL} or \texttt{Eigen} (the latter is used by default in \ambit).

Direct diagonalisation is prohibitively computationally expensive for larger matrices, however. If only 
a few ($<50$) solutions are required, \ambit\ instead solves the CI problem with a Fortran 
implementation of Davidson's 
algorithm \cite{davidson75a} developed by Stathopoulos and Froese Fischer \cite{stathopoulos94a}. 
Davidson's algorithm is an iterative, $\mathcal{O}(N^2)$ algorithm which is rapidly convergent when only
a few extreme (in this case, lowest-lying) eigenvalues and vectors are required. This is commonly the 
case in CI+MBPT calculations. Davidson's algorithm does not modify the matrix elements during the 
solution and only requires matrix-vector multiplications, which can benefit from the internal 
parallelism of linear algebra libraries such as the \texttt{MKL} or \texttt{Eigen}, greatly reducing the
time taken to solve the eigenproblem.

The resulting levels are grouped by total angular momentum $J$ and parity $\pi$, before being ordered by
energy \emph{within} each symmetry. The energy (in both atomic units and cm$^{-1}$), configuration 
percentages and Land\'{e} g-factors are then printed to standard output. A sample of this output for
Cr$^+$ calculations is shown below:

\begin{verbatim}
Solutions for J = 0.5, P = even (N = 204534):
0: -9.8305729    -2157561.36528 /cm
        4s1 3d3 14d1  1.3%
             4s1 3d4  89%
             5s1 3d4  2.8%
    g-factor = 3.331

1: -9.8021109    -2151314.67386 /cm
        4s1 3d3 13d1  1.1%
        4s1 3d3 14d1  1.6%
        4s1 3d3 15d1  1.1%
             4s1 3d4  88%
             5s1 3d4  2%
                 3d5  2%
    g-factor = 0.0028481

 ...

\end{verbatim}

The resulting energy levels are also stored in the binary file \texttt{<ID>_0.levels} in the current
working directory. The software will not generate any stored levels if this file exists prior to starting
the calculation, but will calculate any additional requested levels.

The [CI] section of the input file accepts the following options and flags:

\textbf{LeadingConfigurations} List of all configurations from which to generate many-body 
CSFs for the Hamiltonian matrix. For example, \texttt{CI/LeadingConfigurations='4d5, 4d4 5s1'} will 
build the Hamiltonian matrix by exciting electrons or holes from the two listed configurations 
(4d$^5$ and 4d$^4$5s). The list can contain arbitrarily many leading configurations, but all 
configurations must conserve particle number and must explicitly specify the number of particles in each 
orbital. Holes in an otherwise filled shell (i.e. located below the Fermi level) are denoted by a 
negative occupation number, e.g. \texttt{CI/LeadingConfigurations='3d-1'} contains a hole in the 3d 
shell. (Mandatory.)

\textbf{ExtraConfigurations} List of ``extra'' configurations to be included in the CI matrix. No
electron or holes will be excited from these configurations. Accepts a list of non-relativistic
configurations with the same form as \texttt{CI/LeadingConfigurations}.

\textbf{ElectronExcitations}[2] 
Number of electron excitations to include in the CI procedure. Defaults to 2 if no value is specified.
This option also accepts input as a string of the form \texttt{'1, <Size 1>, 2, <Size 2>, ...'},
which specifies different limits on $pqn$ and $l$ for each electron. For example, the string
\texttt{CI/ElectronExcitations='1,8spdf,2,6spd'} will include all single-excitations up to 8spdf and 
all double-excitations up to 6spd.

\textbf{HoleExcitations}[0] Number of hole excitations to include in the CI procedure. Takes similar
input to \texttt{CI/ElectronExcitations}.

\textbf{EvenParityTwoJ} List of total angular momenta $2J$ of even parity to solve the CI problem for.
For example, \texttt{CI/EvenParityTwoJ='0, 2, 4'} will generate and solve the Hamiltonian matrices for
even parity states with $J = 0, 1, 2$. (At least one of \texttt{CI/EvenParityTwoJ}, 
\texttt{CI/OddParityTwoJ} or \texttt{CI/{-}{-}all-symmetries} must be specfied).

\textbf{OddParityTwoJ} List of total angular momenta $2J$ of odd parity to solve the CI problem for.
For example, \texttt{CI/OddParityTwoJ='1, 3, 5'} will generate and solve the Hamiltonian matrices for
odd parity states with $J = 1/2, 3/2, 5/2$. (At least one of \texttt{CI/EvenParityTwoJ}, 
\texttt{CI/OddParityTwoJ} or \texttt{CI/{-}{-}all-symmetries} must be specfied).

\textbf{{-}{-}all-symmetries} Solves the CI eigenproblem for every available angular momentum of both
parities.

\textbf{NumSolutions}[6] Number of eigenvalues to generate for each symmetry $J^{\pi}$.

\textbf{{-}{-}gfactors} Calculate the Land\'{e} g-factors for each solution (this option is enabled by
default unless \texttt{CI/NumSolutions} $> 50$).

\textbf{{-}{-}no-gfactors} Do not calculate g-factors (since this can take a long time for large $N_{CI}$).

\subsection{[CI/SmallSide] - Emu CI}
Small-side CI is enabled and configured in the sub-section [CI/SmallSide], which accepts
the following options:

\textbf{LeadingConfigurations} List of leading configurations from which to generate the small-side of
the emu CI matrix. Input syntax is as for \texttt{CI/LeadingConfigurations}. (Mandatory if using emu CI.)

\textbf{ElectronExcitations}
Number of electron excitations to include in the ``small side'' of the matrix. This option
accepts the same input format as \texttt{CI/ElectronExcitations}. The \texttt{CI/SmallSide} subsection
is only useful when the ``small side'' of the matrix is smaller than that specified in \texttt{CI}, so
it is important to ensure only the desired configurations are included here.

\textbf{HoleExcitations}[0] - Number of hole excitations to include when building the small-side of the
Hamiltonian matrix.

\textbf{ExtraConfigurations} List of ``extra'' configurations to be included in the small side of the
emu CI matrix. No electron or holes will be excited from these configurations. Accepts a list of
non-relativistic configurations with the same form as \texttt{CI/LeadingConfigurations}.

\subsection{[MBPT]}

MBPT diagrams and corrections are only calculated if appropriate flags are passed to \ambit; 
arguments specified in this section of the input will be ignored by the software unless the 
\texttt{-s[123]} flag has been passed as input.

Once the one- and two-body MBPT integrals have been calculated, they are then stored to disk in the
current working directory as the binary files \texttt{<ID>_0.one.int} and \texttt{<ID>_0.two.int},
respectively. Three-body MBPT integrals are not stored, however, as the number of three-body integrals 
grows very rapidly with the number of orbitals included in MBPT. However each diagram is relatively 
inexpensive given a set of pre-calculated two-body integrals so it is not generally efficient to store 
them in full between calculations. {Three-body MBPT diagrams are only included for Hamiltonian matrix elements where at least one CSF is drawn from the set of leading configurations.
Both one- and two-body MBPT integrals are computationally intensive,
so calculations for these both types of integrals are parallelised by MPI, with the two-body integrals 
being further parallelised by OpenMP.}

Additionally, the MBPT expansion is carried out in the Brillouin-Wigner formalism, in which the energy
denominators in equation \ref{eq:MBPT} depend on the total energy of the valence electrons
\cite{berengut06a, dzuba96a}. While it is straightforward to use the energies obtained from Dirac-Fock 
for the valence energy, there is some computational freedom as to which orbitals to include in the 
denominators (see ref. \cite{berengut06a} for a more detailed discussion of this point). In electron-only
calculations, the software will default to using the orbitals immediately above the Fermi level specified
by the input option \texttt{HF/Configuration}. However, shells which may contain holes in particle-hole
calculations should be included in the energy denominators, in which case it is necessary to explicitly
specify these orbitals via the \texttt{MBPT/EnergyDenomOrbitals} input argument.

Options for MBPT calculations are as follows:

\textbf{Basis} Upper limit on the principal quantum number $n$ and orbital angular momentum $l$ of
virtual orbitals to include in MBPT diagrams. Input string is of the same form as
\texttt{Basis/ValenceBasis}, so \texttt{MBPT/Basis = 30spd20f} will include all s-, p- and d-orbitals
with $n \leq 30$ and f-orbitals with $n \leq 20$ (must be a superset of \texttt{Basis/ValenceBasis}).

\textbf{{-}{-}use-valence} Include valence-valence MBPT diagrams for orbitals above the limit set in
\texttt{Basis/ValenceBasis} and below \texttt{MBPT/Basis}.

\textbf{EnergyDenomOrbitals} Specifies which orbitals/shells to use when calculating the valence energy
in the MBPT diagram energy denominators. Accepts an orbital string as input, so e.g. 
\texttt{MBPT/EnergyDenomOrbitals = 5sp4df} will use the $5s$, $5p$, $4d$ and $4f$ orbitals to calculate
the valence energy.

\subsection{[Transitions] - Matrix elements of external fields}

\ambit\ uses wavefunctions obtained from CI procedure to calculate  matrix elements for multiple
possible operators, {which may be diagonal or non-diagonal (transition) matrix elements.}
In addition to including 
options in [Transitions] in a new (clean) CI+MBPT run, transition matrix elements can also be calculated
using pre-existing solutions from previous calculations, which are stored in the \texttt{<ID>_0.levels}
file.

Matrix elements for the following operators are supported by \ambit:
\begin{itemize}
\item E1, E2, E3 - Electric dipole, quadrupole and octupole operators,
\item M1, M2 - Magnetic dipole and quadrupole operators,
\item HFS1, HFS2 - hyperfine dipole and quadrupole operators,
\end{itemize}
Each of the above can have a different set of requested transitions. In addition, the
software will present either the reduced matrix elements $T = \big< f ||\hat{O}||i\big>$ or line 
strengths $S = |T|^2$ for each operator $\hat{O}$, but not both. If no options are specified the 
line-strengths will be calculated.

There are two possible ways of requesting transitions: either a set of individual transitions,
or all transitions involving levels below a specified energy threshold can be targeted. Individual 
transitions must be requested according to the scheme \texttt{<2J><parity>:<N>}, where 2J is an integer 
equal to twice the total angular momentum, parity is either \texttt{e} or \texttt{o} (even or odd) and 
N is the integer used to order the level in the CI output. For example, \texttt{0e:0 -> 2o:1}
represents the transition from the lowest energy state with even parity and angular momentum $J = 0$ to
the second lowest-energy state with odd parity and angular momentum $J = 1$). The transitions requested 
need not necessarily be the same for each multipole operator.

Options for each operator are specified in separate subsections, such as
\texttt{Transitions/E1/{-}{-}reduced-elements}. Each subsection accepts the same set of possible 
arguments, but the arguments need not be the same for each requested operator. The arguments accepted 
by each subsection are as follows:

\textbf{MatrixElements} List of specific transition matrix elements to calculate, e.g.
\mbox{\texttt{Transitions/M1/MatrixElements = '1e:0 -> 1e:1, 3o:2 -> 3o:4'}}. Diagonal matrix elements 
for operators of, e.g. hyperfine structure, may also be given in the form 
\mbox{\texttt{Transitions/HFS1/MatrixElements = '2e:0, 4e:0'}}.

\textbf{AllBelow} [0.0] Calculates all matrix elements between states with energy less than the
argument.

\textbf{Frequency} Frequency of the external field (in atomic units). Only used with EJ and MJ
operators. If not specified the default is to use the Dirac-Fock transition frequency; RPA must then be
recalculated for each transition.

\textbf{{-}{-}reduced-elements} Calculate the reduced matrix elements $T$ rather than the line 
strengths $S = |T|^2$.

\textbf{{-}{-}print-integrals} Prints the raw value of the nonzero reduced one-body matrix elements 
$\big< a|| \hat{O} ||b \big>$ for each pair of valence orbitals $a$ and $b$.

\textbf{{-}{-}rpa} Include frequency-dependent RPA corrections to the transition matrix elements.

\textbf{RPA/{-}{-}no-negative-states} Exclude basis states in the Dirac sea (i.e. negative energy
states) from RPA corrections.

\subsection{Parallelism and performance}

\ambit\ is highly optimised to take advantage of modern high-performance computing hardware and makes
extensive use of modern parallel programming techniques. We employ a hybrid model of parallelism, which
uses MPI to divide the workload between computational resources (such as nodes on an HPC cluster) and
OpenMP threads to further subdivide the workload between cores on a single resource. This approach 
allows us a
fine-grained control over the workload distribution to match the hardware and network topologies of a
particular system, providing greater per-node performance than a pure MPI approach. More importantly,
part of the CI matrix must be duplicated between MPI processes, so this hybrid MPI+OpenMP approach is
significantly more memory-efficient than pure MPI.

Creating and diagonalising the CI matrix is by far the most computationally expensive part of most 
CI+MBPT calculations, so this part of the software has been extensively optimised for both run-time and
memory usage. First, the CI matrix is divided up into ``chunks'', each of which contains the CSFs from
four relativistic configurations. These chunks are then distributed approximately evenly between MPI 
processes, each of which then calculates the elements of the chunk in parallel via OpenMP. 
Additionally, the matrix is sorted 
such that the configurations with the most projections appear first. These rows are the most expensive 
to calculate, so this ensures that the OpenMP workload for each process is well balanced.

Once the CI matrix chunks have been populated, the matrix is then solved via Davidson's method
\cite{davidson75a, stathopoulos94a}, which distributes the necessary matrix-vector multiplications 
approximately evenly among MPI processes. These matrix-vector multiplications are trivially 
parallelisable, so we exploit internal OpenMP parallelism provided by either \texttt{Eigen} or Intel
\texttt{MKL} to further parallelise each MPI process's workload. This results in a dramatic speedup in
matrix diagonalisation: even CI matrices with $10^5$ CSFs on a side can usually be diagonalised in a
few minutes on a mid-sized HPC cluster.

In addition to CI, we also parallelise generating the Slater (CI) and MBPT integrals, which can dominate
the calculation run-time for one and two valence-electron systems with a very large basis. Land\'{e}
g-factors and transition matrix elements are also parallelised.

Finally, it is important to ensure that the parallel workload is divided appropriately for the topology
of the target machine; incorrect choice of parallelism can severely impact performance. 
Usually, it is best to have spawn MPI process per node and as many OpenMP threads
as there are cores per node. This is not the default behaviour for many MPI implementations
(including \texttt{OpenMPI} and \texttt{IntelMPI}), so
it must be explicitly requested when running \ambit. Note that some hardware architectures, such as 
those with non-uniform memory access (NUMA) nodes may be better served by other distributions of threads
and processes.

\section{Example calculation: Cr$^+$}
In this section, we present results of two full CI+MBPT calculations for the spectrum of Cr$^+$ to
demonstrate the usage and features of \ambit. All input
and output files for this calculation are included as appendices in the folder \texttt{CrIIExample}.
Cr$^+$ is a five-electron system which requires a large CI basis and both core- and 
valence-MBPT corrections in order to produce an accurate spectrum, making it a good test-case for the
theoretical and computational techniques included in \ambit. 

The two calculations presented here represent two computational regimes: a relatively small-scale
calculation which requires only a single workstation or compute server, as well as a much larger scale 
calculation taking better advantage of modern HPC architecture. Both sets of calculations utilise emu CI
and demonstrate that this technique allows for significantly higher accuracy than previous, standard
CI+MBPT calculations \cite{berengut11b}.

Energy levels from these calculations are shown in tables \ref{tab:CrII_spectra} and \ref{tab:EmuCI}. 
Table \ref{tab:CrII_spectra} additionally details the contributions to the overall energy from CI, one-,
two-, and three-body core-valence MBPT. These results are also compared with experimental spectra from 
\cite{nistasd54}.

\subsection{Dirac-Fock and B-spline basis}

All of the Cr$^+$ calculations presented in this section were undertaken using a $V^N$ Dirac-Fock
potential, including all 3d$^5$ valence electrons in the DF potential. This choice of
potential necessarily introduces subtraction diagrams to MBPT, but these did not significantly degrade
the accuracy of the calculations due to the large-CI basis employed throughout. We generated a frozen
core consisting of filled 1s, 2s, 2p, 3s and 3p shells and included 3d valence orbitals above the Fermi
level, as described in section \ref{sec:comp_HF}. Additionally, we generate all single-particle 
B-Spline orbitals up to 16spdf for pure CI and 30spdfg for MBPT using the $V^N$ Dirac-Fock operator.

The choice of $V^N$ potential produces 3d orbitals which are spectroscopic for the 3d$^5$
configuration, but which are less accurate at treating the 3d$^4$ 4s configurations. Forming 3d orbitals
with five electrons results in a Dirac-Fock wavefunction which is less tightly bound to the nucleus and
thus has higher energy than the corresponding 3d$^4$ DF orbital. Although this means our calculations
tend to underestimate the energy of the levels with 3d$^4$ 4s configuration, we find that use of the 
$V^{N-1}$ potential is much more slowly convergent for the ground-state 3d$^5$ wavefunction, resulting 
in significantly reduced accuracy compared to our choice of $V^N$ potential.

\subsection{Small-scale calculation}
Our small-scale calculation targets even-parity states with $J = 1/2, 3/2, 5/2, 7/2, 9/2$ and forms the 
CI configuration electron excitations from the 3d$^5$, 
3d$^4$4s and 3d$^4$4d leading configurations. Additionally, we employ the emu CI technique in section
\ref{sec:SmallSide} with a ``large-side'' formed from single- and double-excitations up to 16spdf, 
and a ``small-side'' formed from single- and double-excitations up to 5spdf. This choice of CI basis 
results in CI matrices with $\sim 10^6 \times 10^4$ nonzero elements, the largest of which 
consisted of $440704 \times 9449$ CSFs for $J = 5/2$.

Table \ref{tab:CrII_spectra} shows a break-down of the effects of one- two- and three-body
core-valence MBPT corrections with a basis of 30spdfg.
As previously stated, these calculations use the $V^N$ potential, which necessarily introduces large 
subtraction diagrams into the MBPT expansion as outlined in section \ref{sec:CI+MBPT}.
As can be seen in the $\Sigma^{(1)}$ column of table \ref{tab:CrII_spectra}, including the one-body
subtraction diagrams significantly reduces the accuracy of the resulting spectra; the calculations no
longer correctly identify even the $3d^5$ ground-state. However, including the two- and three-body MBPT
corrections, shown in the $\Sigma^{(2)}$ and $\Sigma^{(3)}$ columns in table \ref{tab:CrII_spectra},
restores and improves the accuracy of the calculations due to the partial cancellation of subtraction
diagrams by two- and three-body terms \cite{berengut11b}. 

Both CI-only and CI+MBPT calculations required less than 40GB of memory and less than 1 hour of compute 
time when distributed between 16 cores -- well within the capabilities of a high-performance workstation
or a single node of a typical HPC cluster. This calculation does not approach saturation of the CI 
basis since it is constrained to require only relatively modest computational resources. However once 
all core-valence MBPT diagrams have been included, the CI+MBPT energy levels shown in table 
\ref{tab:CrII_spectra} have an average error of less than 5\%.

\begin{table}
\caption{Energy levels of Cr$^+$ (in cm$^{-1}$) for small-scale emu CI calculation. The ``large-side''
of the matrix contains single- and double-excitations up to 16spdf, while the ``small-side'' contains
single- and double-excitations up to 5spdf. The first and second columns
give the nonrelativistic configuration and approximate $LS$-coupling term for each calculated level. The
CI column gives the (excitation) energies as calculated using a pure-CI approach, while the
$\Sigma^{(1,2,3)}$ columns give the one-, two- and three-body MBPT corrections. Finally, the total 
energy and experimental energies obtained from ref. \cite{nistasd54} are presented along with the
percentage differences between experiment and our calculations.}
\label{tab:CrII_spectra}
\begin{tabular}{l l l l l l l l l}
\hline
Configuration    &Term    &CI    &$\Sigma^{(1)}$    &$\Sigma^{(2)}$    &$\Sigma^{(3)}$    &Total
&Expt    &$\Delta_{\mathrm{Expt.}}$ 
(\%)\\
\hline
\hline
3d$^5$      &$^6S_{5/2}$    &0    &0    &0    &0    &0    &0    &--\\
3d$^4$ 4s   &$^6D_{1/2}$    &14038  &-15449    &12116    &2038    &12742    &11962    &-6.5\\
3d$^4$ 4s   &$^6D_{3/2}$    &14087  &-15432    &12102    &2039    &12795    &12033    &-6.3\\
3d$^4$ 4s   &$^6D_{5/2}$    &14186  &-15420    &12087    &2040    &12893    &12148    &-6.1\\
3d$^4$ 4s   &$^6D_{7/2}$    &14446  &-15488    &12129    &2037    &13124    &12304    &-6.7\\
3d$^4$ 4s   &$^6D_{9/2}$    &14813  &-15619    &12218    &2035    &13448    &12496    &-7.6\\
3d$^4$ 4s   &$^4D_{1/2}$    &20285  &-14470    &11807    &2319    &19941    &19528    &-2.1\\
3d$^4$ 4s   &$^4D_{3/2}$    &20415  &-14478    &11801    &2321    &20060    &19631    &-2.2\\
3d$^4$ 4s   &$^4D_{5/2}$    &20707  &-21512    &18797    &2323    &20315    &19798    &-2.6\\
3d$^4$ 4s   &$^4D_{7/2}$    &21104  &-22146    &19389    &2323    &20671    &20024    &-3.2\\
\hline
\end{tabular}
\end{table}

\subsection{Large-scale calculation}

The large-scale calculation targets the same $J^\pi$ symmetries as the small-scale calculation, but 
employs a more sophisticated method of constructing the CI matrix. Here, we form the large-side from all
single- and double- excitations from the 3d$^5$, 3d$^4$4s and 3d$^4$4d leading configurations up to
15spdf. We then form the small-side from all single-excitations up to 15spdf and single- and
double-excitations up to 5spdf. This approach more accurately captures the important configurations in
the CI expansion at the expense of significantly larger matrix sizes. {See listing} 
\ref{lst:Cr+Input} {for the full input file used to generate the emu CI-only spectrum.} 
The largest matrix ($J = 5/2$) has $374944 \times 109779$ non-zero elements and requires approximately
550GB of memory, which, while large, is still much less than full-CI and is still within the
capabilities of modern HPC clusters.

Table \ref{tab:EmuCI} shows the calculated spectra from these large-scale CI-only calculations, as well
as the effects of core-valence MBPT with a basis of 30spdfg. {The CI-only calculations in table}
\ref{tab:EmuCI} give excellent
agreement with experiment, with average errors at the $1\%$ level. However, unlike the smaller
calculations (and CI+MBPT calculations from other codes), the inclusion of MBPT slightly degrades the 
calculation's accuracy. Further increasing the size of the MBPT basis via the \textit{MBPT/Basis} does not increase the accuracy beyond that of the CI-only
calculation. This behaviour suggests that CI+MBPT has qualitatively different convergence at
regimes close to saturation of the CI expansion for open-shell atoms, so care must be taken when using
large-scale calculations for these systems.

\begin{table}
\caption{Energy levels of Cr$^+$ (in cm$^{-1}$) for large-scale emu CI calculation. The ``large-side''
of the matrix contains single- and double-excitations up to 16spdf, while the ``small-side'' contains
single-excitations up to 15spdf and single- and double-double-excitations up to 5spdf. The first and 
second columns give the nonrelativistic configuration and approximate $LS$-coupling term for each 
calculated level. The $E_{\mathrm{CI}}$ and $\Delta_{\mathrm{CI}}$ columns give the (excitation) 
energies and difference from experimental values as calculated using a pure-CI approach. The
$E_{\mathrm{MBPT}}$ and $\Delta_{\mathrm{MBPT}}$ columns show the same comparison for CI+MBPT. The
experimental values \cite{nistasd54} are shown under $E_{\mathrm{Expt}}$.}
\label{tab:EmuCI}
\begin{tabular}{l l l l l l l}
\hline
Configuration    &Term    &E$_{\mathrm{CI}}$    &$\Delta_{\mathrm{CI}}$ (\%) & E$_{\mathrm{CI+MBPT}}$  &
$\Delta_{\mathrm{CI+MBPT}}$ (\%) & $E_{\mathrm{Expt}}$\\
\hline
\hline
3d$^5$    &$6S_{5/2}$    &0    &--    &0    &--  &0\\
3d$^4$ 4s    &$6D_{1/2}$    &11956    &0.05    &11237    &6.1    &11962\\
3d$^4$ 4s    &$6D_{3/2}$    &12072    &-0.3    &11341    &5.7    &12033\\
3d$^4$ 4s    &$6D_{5/2}$    &12265    &-1.0    &11512    &5.2    &12148\\
3d$^4$ 4s    &$6D_{7/2}$    &12531    &-1.8    &11750    &4.5    &12304\\
3d$^4$ 4s    &$6D_{9/2}$    &12867    &-3.0    &12048    &3.6    &12496\\
3d$^4$ 4s    &$4D_{1/2}$    &19441    &0.4    &19605    &-0.4    &19528\\
3d$^4$ 4s    &$4D_{3/2}$    &19624    &0.04    &19771    &-0.7   &19631\\
3d$^4$ 4s    &$4D_{5/2}$    &19921    &-0.6    &20043    &-1.2   &19798\\
3d$^4$ 4s    &$4D_{7/2}$    &20320    &1.5    &20398    &1.9   &20024\\
\hline
\end{tabular}
\end{table}

\begin{lstfloat}
\caption{AMBiT input file used to generate Cr$^+$ emu CI spectra in table \ref{tab:EmuCI}.}
\begin{verbatim}
ID=CrII                                                                        
Z=24                                                                           

[Lattice]                                                                      
NumPoints=1000                                                         
StartPoint=1.0e-6                                                      
EndPoint=60.0                                                          

[HF]                                                                           
N=23                                                                        
Configuration='1s2 2s2 2p6 3s2 3p6 : 3d5'

[Basis]                                                                        
--bspline-basis                                                          
ValenceBasis=15spdf                                                      
FrozenCore=3sp                                                           
BSpline/Rmax=60.0                                                        

[CI]                                                                           
LeadingConfigurations='3d5, 3d4 4s1, 3d4 4p1'                                 
ElectronExcitations=2                                                       
HoleExcitations=0                                                           
EvenParityTwoJ='1, 3, 5, 7, 9, 11'                                            
NumSolutions=6                                                              
                            
[CI/SmallSide]                                                                 
LeadingConfigurations='3d5, 3d4 4s1, 3d4 4p1'                       
ElectronExcitations='1,15spdf, 2,5spdf' 
\end{verbatim}
\label{lst:Cr+Input}
\end{lstfloat}

\section{Acknowledgements}
We thank A. J. Geddes for feedback and testing for \ambit\, as well as reviewing this manuscript.
The work of EVK was supported by the Australian Government Research Training Program scholarship. The
work of JCB was partially funded by the Australian Research Council grant DE120100399.

\FloatBarrier

\bibliographystyle{elsarticle-num}
\bibliography{atoms}

\end{document}